\documentclass[showpacs,aps,twocolumn,floatfix,prl,longbibliography]{revtex4-2}
\usepackage{bm,amsmath,amssymb, mathrsfs}
\usepackage{bbm} 
\usepackage{physics}
\usepackage{commath}
\usepackage{braket} 
\usepackage{graphicx,bm}
\usepackage{verbatim}
\usepackage[dvipsnames]{xcolor}
\usepackage{soul}
\usepackage[separate-uncertainty = true]{siunitx}
\usepackage[colorlinks]{hyperref}
\AtBeginDocument{%
    \hypersetup{    
        linkcolor=blue,
        filecolor=magenta,      
        urlcolor=blue,
        citecolor=blue,
        colorlinks=true,
     }
}

\newcommand{\eu}{\mathrm{e}}
\newcommand{\imi}{\mathrm{i}}

\usepackage[T1]{fontenc}
\usepackage[utf8]{inputenc}

\begin{document}

\title{Polarization-dependent topology in quantum emitter chains}

\author{Jonathan Sturm}
\email{jonathan.sturm@uni-wuerzburg.de}

\author{Adriana P\'alffy}
\email{adriana.palffy-buss@uni-wuerzburg.de}

\affiliation{Julius-Maximilians-Universit{\"a}t W{\"u}rzburg, Institute for Theoretical Physics and Astrophysics and W{\"u}rzburg-Dresden Cluster of Excellence ct.qmat, Am Hubland, 97074 W{\"u}rzburg, Germany}

\date{\today}

\begin{abstract}
The role of polarization in the topology of quantum emitter chains is investigated theoretically, whereby ``polarization'' refers to the transition dipole moments of the emitters.
We show that, if the chain is zigzag-shaped, different topological phases can be realized by adjusting the polarization direction. 
It turns out that long-range dipole-dipole couplings weaken the bulk-boundary correspondence, but give rise to higher-order topological phases with four observable edge modes.
We also demonstrate how the polarization orientation can be used to define an additional dimension and simulate a synthetic Chern insulator. 
Our findings open up a way to actively switch between various topological phases within a single arrangement of quantum emitters.
\end{abstract}

\maketitle

\textit{Introduction.} Topological states of matter have attracted enormous interest over the past decades, as they can provide robust boundary states protected by global bulk properties.
While topological phenomena were originally discovered in electronic platforms, like quantum Hall systems \cite{vonKlitzing1980, Haldane1988}, topological insulators \cite{Koenig2007, Qi2011}, or topological superconductors \cite{Kitaev2001, Fu2008}, the fundamental concepts of Berry phase \cite{Berry1984} and symmetry classes \cite{Ryu2010}  have by now been successfully transferred to a large variety of platforms, including photonic \cite{Haldane2008, Raghu2008}, acoustic \cite{Yang2015}, and mechanic systems \cite{Huber2016}. 
Moreover, Perczel \textit{et al.} have successfully applied these concepts to quantum optics by showing that a hexagonal array of quantum emitters can develop protected chiral edge modes, similar to a Chern insulator \cite{Perczel2017}.

Quantum emitters, modeled as two- or multilevel systems, can couple via dipole-dipole interactions, where an excitation can jump from one emitter to another by exchange of virtual photons \cite{Lehmberg1970, Lehmberg1970II, Agarwal_book, Garcia2017, Reitz2022}.
Moreover, the emitters dissipate collectively, leading to either an increased decay rate (superradiance) \cite{Dicke1954, Gross1982, Devoe1996} or suppressed emission (subradiance) \cite{Devoe1996, Guerin2016}.
Quantum emitter arrays have been successfully realized in a plethora of platforms, such as cold atoms \cite{Nascimbene2015, Du2024}, optical tweezer arrays \cite{Endres2016, Barredo2016, Leseleuc2019}, nitrogen-vacancy centers \cite{Dolde2013, Juan2017}, quantum dots \cite{Brandes2005}, or superconducting qubits \cite{Nissen2013, Mlynek2014}.
So far, topological edge modes in quantum emitter lattices have been predicted for honeycomb lattices \cite{Perczel2017, Bettles2017, Weber2018, Wulles2024}, square lattices \cite{Bettles2017}, Hofstadter systems \cite{Zhang2019, Wang2021}, and dimerized chains \cite{Wang2018, McDonnel2022, Svendsen2024, Wang2024}, where for the latter it has been shown that the topological phase can lead to single-photon emission \cite{Wang2024}.
In these studies, topology was induced by either specific emitter positioning \cite{Wang2018, McDonnel2022, Svendsen2024, Wang2021} or external fields \cite{Perczel2017, Bettles2017, Weber2018, Wulles2024, Zhang2019}. 

This Letter focuses on a quantity that has not attracted much attention so far in the context of quantum emitters: the orientation of the transition dipole moments of the emitters, also referred to as the polarization of the array.
While, for instance, Ref.~\cite{Wang2018} has pointed out that the polarization has no influence on the topology of a linear quantum emitter chain, we will show that if the chain geometry is altered from a linear arrangement to a zigzag shape, bulk topology can be controlled using in-plane polarization. 
This renders possible the realization of various topological phases within the same emitter geometry via controllable dipole orientations.
While, to our knowledge, this is a novel concept in the context of quantum emitters, analogs of this approach have been implemented in plasmonic chains \cite{Poddubny2014, Sinev2015, Slobozhanyuk2015, Kruk2017, Kruk2019, Zhang2021, Moritake2022, Wu2024} and polariton systems \cite{Solnyshkov2016, StJean2017}.

We also study how the long-range dipole-dipole couplings between emitters weaken the established bulk-boundary correspondence \cite{Bettles2017, Perez2019}, while at the same time allowing for higher-order topological phases. 
For emitters with freely rotatable dipole moments, we show how the polarization can be used to introduce an additional synthetic dimension and simulate a quantum optical Chern insulator. 
Our findings pave the way toward dynamical control of topological phases in quantum emitter chains.

\textit{Model.}
\begin{figure*}
    \centering
    \includegraphics[width=1\linewidth]{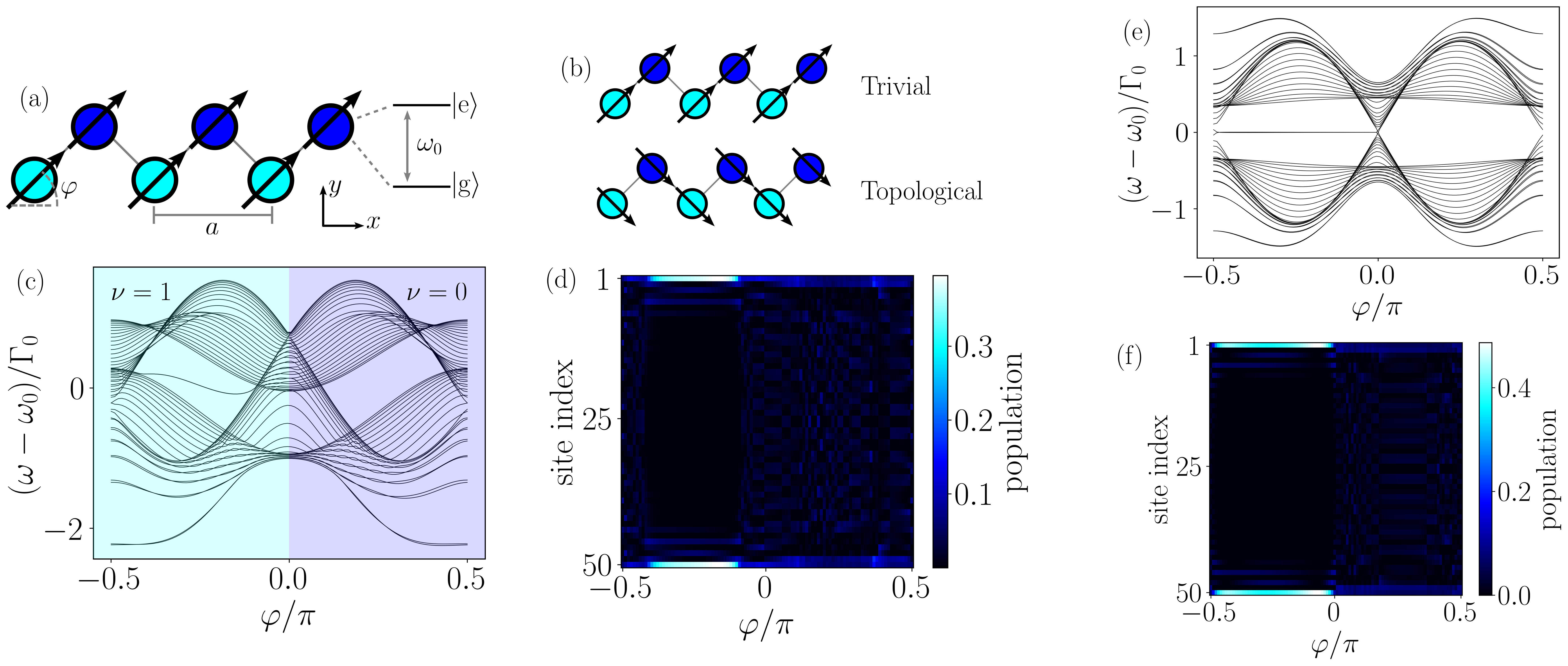}
    \caption{(a) A zigzag chain with lattice constant $a$ (colors indicating the two sublattices) of two-level systems coupled via dipole-dipole interactions,  with ground (excited) state $\ket{\mathrm{g}}$  ($\ket{\mathrm{e}}$), transition frequency $\omega_0$, and  angle $\varphi$ of the dipole moments relative to the $x$ axis. 
    (b) Different orientations of the dipole moments enable different topological phases.
    (c) Spectrum of the Hamiltonian in Eq.~\eqref{eq draft Hamiltonian}.
    Shaded areas indicate different values of the winding number $\nu$.
    (d) Populations $|\braket{\psi|i}|^2$ as a function of site index $i$ and $\varphi$.
    Hereby, for each $\varphi$ we consider the state $\ket{\psi}$ with the highest amplitude on the edge $\ket{1}$.
    (e) Spectrum of the Hamiltonian when neglecting intrasublattice couplings.
    (f) Populations computed as in panel~(d) for the same reduced Hamiltonian. 
    All plots used $N=50$ atoms and $a=0.3\lambda_0$.
    }
    \label{draft fig 1}
\end{figure*}
We consider an array of $N$ identical two-level systems (``atoms'') with ground and excited states $\ket{\mathrm{g}_i}$ and $\ket{\mathrm{e}_i}$, with $i=1,...,N$ and transition frequency $\omega_0$.
The atoms are arranged in a zigzag geometry in free space with positions $\Vec{r}_{2j-1} = (j-1)\vec{a}$ and $\Vec{r}_{2j} = \frac{a}{2}(0, \ 1, \ 0)^\mathrm{T} + (j-1)\vec{a}$, respectively, with $j=1,...,N/2$ and fundamental lattice vector $\vec{a}=a(1, \ 0, \ 0)^\mathrm{T}$ with lattice constant $a>0$, as shown in Fig.~\ref{draft fig 1}(a).
The atoms interact via resonant dipole-dipole couplings and show incoherent collective decay such that the time evolution of a state described by a density matrix $\varrho$ is given by the master equation $\Dot{\varrho} = -\imi [H,\varrho] + \mathscr{L}[\varrho]$ \cite{Dung2002, Garcia2017, Garcia2017_A, Reitz2022}, where the Hamiltonian $H$ is given by 
\begin{equation}\label{eq draft Hamiltonian}
    H = \sum_{i=1}^N \omega_0 \sigma_i^\dagger \sigma_i 
    + \sum_{i\neq j} \Omega_{ij} \sigma_i^\dagger \sigma_j,
\end{equation}
while the Lindblad operator reads
\begin{equation}\label{eq draft lindblad}
    \mathscr{L}[\varrho] = \sum_{i,j=1}^N \frac{\Gamma_{ij}}{2} (2\sigma_i \varrho \sigma_j^\dagger - \sigma_i^\dagger \sigma_j \varrho - \varrho \sigma_i^\dagger \sigma_j).
\end{equation}
Here, $\sigma_i=\ket{\mathrm{g}_i}\bra{\mathrm{e}_i}$ are the atomic transition operators and $\hbar=1$.
The dipole-dipole couplings $\Omega_{ij}$ and the collective decay rates $\Gamma_{ij}$ are given by
\begin{equation}\label{eq draft Omegaij}
    \Omega_{ij} -\imi\frac{\Gamma_{ij}}{2}
    = -\Gamma_0 \frac{3\pi}{k_0} \ \hat{p}_i \cdot 
    G(\Vec{r}_i-\Vec{r}_j, \omega_0) \cdot \hat{p}_j,  
\end{equation}
where $\hat{p}_i=\Vec{p}_i/|\Vec{p}_i| = (\cos(\varphi), \sin(\varphi), 0)^\mathrm{T}$ are the normalized atomic transition dipole moments, $\Gamma_0$ is the single atom decay rate, and $k_0 = 2\pi/\lambda_0 =\omega_0/c$.
Moreover, $G$ is the free-space electromagnetic Green's function:
\begin{align}
    G(\Vec{r},\omega_0) = \frac{\eu^{\imi k_0 r}}{4\pi k_0^2 r^3}
    \Big(
    & (k_0^2r^2 + \imi k_0 r -1)\mathbbm{1} \nonumber  \\
    &  
    + (-k_0^2r^2 - 3\imi k_0r + 3) \mathrm{pr}_{\Vec{r}}
    \Big),
\end{align}
where $r=|\Vec{r}|$ and $\mathrm{pr}_{\Vec{r}} = \Vec{r}\otimes\Vec{r}/r^2$ is the projector on $\Vec{r}$.
Note that since this Letter focuses on studying topological phases, which are a static property of the Hamiltonian, we are mostly working with the Hamiltonian given in Eq.~\eqref{eq draft Hamiltonian}. 
Dissipation becomes relevant when we realize a synthetic Chern insulator, which is discussed in the last section of this Letter. \\

\textit{Polarization-dependent topology.}
The bipartite nature of our chain resembles the Su-Schrieffer-Heeger (SSH) model \cite{Su1979}, which consists of two sublattices with nearest-neighbor interatomic couplings $t$ (couples atoms within a unit cell) and $t'$ (couples atoms between two unit cells) \cite{Su1979, Reitz2022, Asboth_book}.
Depending on the ratio of these couplings, the chain takes two distinct topological phases characterized by a bulk topological invariant, the winding number, and the presence or absence of protected localized boundary modes \cite{Asboth_book}. 
In our system, the dipole-dipole couplings $\Omega_{ij}$ play the roles of $t$ and $t'$ and depend on two parameters, the distance $\Vec{r}_i-\Vec{r}_j$ and the orientation of the dipole moments $\hat{p}_i$. However, in our case, the interactions are long-ranged and extend beyond nearest-neighbor coupling.
There are several examples in the literature that treat quantum optical implementations of the SSH model in which the topological phases are determined by the interatomic distances \cite{Wang2018, McDonnel2022}.
It has also been pointed out in Ref.~\cite{Wang2018} that, in a linear chain, the polarization does not affect the topology.

This is not the case for the zigzag chain.
Since the nearest-neighbor couplings $\Omega_{i,i+1}$ are minimal (maximal) if the dipole moments are perpendicular (parallel) to the atomic distance vector $\Vec{r}_i - \Vec{r}_{i+1}$, we expect the chain to be in a topologically trivial (nontrivial) phase for $\varphi=\pi/4$ ($\varphi=-\pi/4$) [see Fig.~\ref{draft fig 1}(b)].
Similar considerations have been demonstrated theoretically and experimentally in the context of coupled polariton micropillars in Refs.~\cite{Solnyshkov2016, StJean2017}, as well as in plasmonic chains \cite{Poddubny2014, Sinev2015, Slobozhanyuk2015, Kruk2017, Kruk2019, Zhang2021, Moritake2022, Wu2024}.
Here, we study how polarization can be used to switch the topological phase within the same quantum emitter zigzag chain. 
We remark that the intuitive picture shown in Fig.~\ref{draft fig 1}(b) is reversed if $a>\lambda_0/2$ due to the complex nonmonotonous $r$-dependence of the Green's function.
However, since $a<\lambda_0/2$ is required for interesting collective effects like subradiance \cite{Garcia2017}, we will always restrict our calculations to this regime.

By restricting the Hamiltonian in Eq.~\eqref{eq draft Hamiltonian} to the single-excitation subspace defined by the basis $\{\ket{i}=\sigma_i^\dagger\ket{\mathrm{GS}}\,|\, i=1,...,N\}$ (where $\ket{\mathrm{GS}}$ is the collective ground state), we can calculate the spectrum as a function of the polarization angle from the Schrödinger equation $(H-\omega(\varphi))\ket{\psi}=0$, which yields the band structure in Fig.~\ref{draft fig 1}(c).
The plot shows that for $\varphi<0$ two degenerate midgap states near $\omega-\omega_0=0$ appear, which do not exist for $\varphi>0$.
These midgap states correspond to localized edge states protected by the bulk topological invariant $\nu$ known as the winding number, whose values are indicated by the different background colors in Fig.~\ref{draft fig 1}(c).
The winding number can be computed from the Bloch Hamiltonian $\mathcal{H}(k)=d_0(k)\sigma_0 + \Vec{d}(k)\cdot\Vec{\sigma}$ (where $\Vec{d}: \mathrm{BZ}\to\mathbb{R}^3$ is the Bloch vector mapping on the Brillouin zone BZ and $\Vec{\sigma}$ is the vector of Pauli matrices),
\begin{equation}
    \nu = \frac{1}{2\pi} \int_\mathrm{BZ} \frac{d_x \partial_k d_y - d_y \partial_k d_x}{d_x^2 + d_y^2} \dd{k},
\end{equation}
and describes the winding of $(d_x, d_y)$ around $0$ in the $d_x$-$d_y$ plane \cite{Asboth_book}.
The Bloch Hamiltonian itself can be found from Eq.~\eqref{eq draft Hamiltonian} by Fourier-transforming $\sigma_i = \frac{1}{\sqrt{N/2}} \sum_k \eu^{-\imi ak i} \Tilde{\sigma}_k$.
There is an important difference between our Bloch Hamiltonian and the Bloch Hamiltonian of the standard SSH model.
The long-range nature of $\Omega_{ij}$ includes couplings between atoms of the same sublattice, leading to the $d_0$ term in $\mathcal{H}(k)$ proportional to the unit matrix $\sigma_0$. 
This term leads to a formal breaking of the chiral symmetry condition $\sigma_z \mathcal{H}(k) \sigma_z = -\mathcal{H}(k)$ since $\sigma_z\sigma_0\sigma_z=\sigma_0\neq-\sigma_0$.
Therefore, it is often said that intrasublattice couplings move the Hamiltonian from symmetry class BDI to class AI \cite{McDonnel2022}, which generally does not provide nontrivial topological phases in one dimension \cite{Ryu2010}. 
As it has been pointed out in Refs.~\cite{Pocock2018, Wang2018}, this is a somewhat misleading statement since the winding number only depends on the Bloch vector $\Vec{d}(k)$, and therefore, the $d_0$ term does not affect the bulk topology.
That is why we find a non-trivial winding number for $\varphi<0$.

Even though intrasublattice couplings do not alter the bulk topological invariant, they still have an impact on the edge states predicted by bulk-boundary correspondence. 
Figure~\ref{draft fig 1}(d) shows for each $\varphi$ the populations $|\braket{\psi|i}|^2$ for the state $\ket{\psi}$ with the highest amplitude on the edge $\ket{1}$.
Although the winding number is equal to 1 for $\varphi<0$, localized edge modes are only appearing in the range of $-0.4\lesssim \varphi/\pi \lesssim -0.1$.
The intrasublattice couplings lead to a spectrum that is not particle-hole symmetric, 
i.e., symmetric with respect to $\omega_0$,
as it becomes evident in Fig.~\ref{draft fig 1}(c).
In vicinity of the phase transition, the edge state bands submerge in the bulk bands, causing the edge states to disappear \cite{Perez2019}.
Therefore, around the phase transition the topological invariant fails to reliably predict the presence of edge modes, violating bulk-boundary correspondence.
To further clarify this, we show in Fig.~\ref{draft fig 1}(e) the spectrum calculated by artificially removing the intrasublattice couplings from the Hamiltonian.
The spectrum becomes particle-hole symmetric with two states at $\omega-\omega_0=0$ for $\varphi<0$ and Dirac cones at $\varphi=0$ and $\varphi=\pm\pi/2$.
For the populations, we see in Fig.~\ref{draft fig 1}(f) that the state is well localized at the edges for $\varphi<0$ with a sharp transition at $\varphi=0$.
This proves that the intra-sublattice couplings, despite only formally breaking chiral symmetry and conserving the topological invariant, weaken the bulk-boundary correspondence around the phase transition.

\textit{Influence of long-range couplings.}
\begin{figure}
    \centering
    \includegraphics[width=1\linewidth]{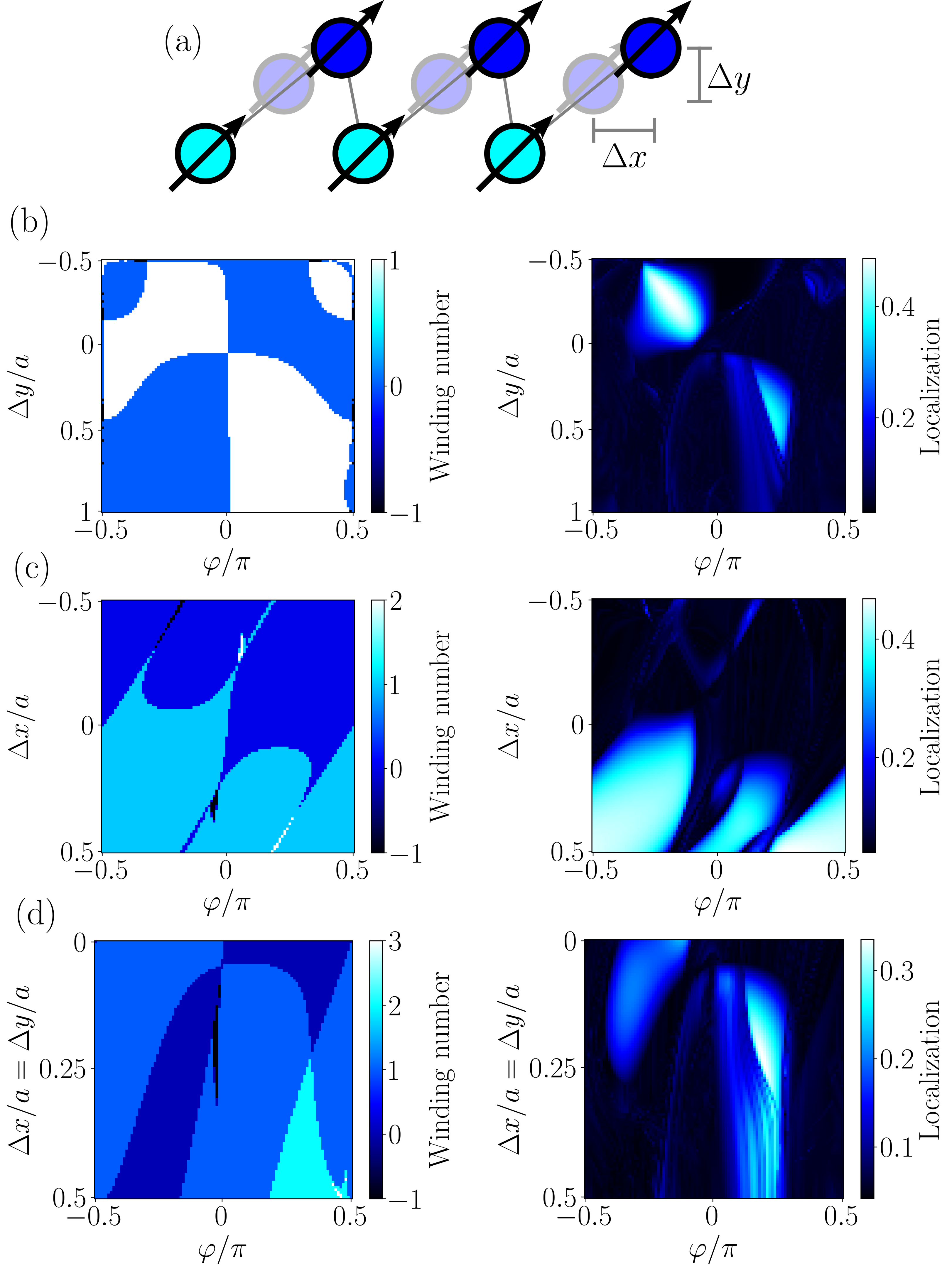}
    \caption{(a) System with shifted sublattice (the original position indicated by the desaturated points).
    (b)-(d) Winding number and localization [cf. Eq.~\eqref{eq draft loc}] as functions of polarization angle $\varphi$ and the sublattice shifts. 
    We used $a=0.35\lambda_0$ and $N=50$ in all plots.
    }
    \label{draft fig 2}
\end{figure}
We have seen that the existence of a nontrivial bulk topological invariant does not necessarily imply existence of localized edge modes. 
Such breakdown of the conventional bulk-boundary correspondence has been observed in other systems with long-range couplings before \cite{Perez2019, Pocock2019}.
While Ref.~\cite{Pocock2019} studied this breakdown based on the coalescence of bulk and edge energy bands, we propose a more direct method by introducing a quantity that reliably predicts the presence of localized modes.
Starting from the definition of the inverse participation ratio (IPR) of a state $\ket{\psi}$,  $ \mathrm{IPR}(\ket{\psi}) = \sum_{i=1}^N |\braket{\psi|i}|^4$ \cite{Goetschy2011}, 
we define the maximal localization of states as
\begin{equation}\label{eq draft loc}
    \mathrm{Loc} = \max_{i=1,...,N} \mathrm{IPR}(\ket{\psi_i}),
\end{equation}
where $\ket{\psi_i}$ are the eigenstates of the real-space Hamiltonian in Eq.~\eqref{eq draft Hamiltonian}.
The IPR takes values between $\frac{1}{N}$ and 1, where 1 means the state $\ket{\psi}$ is completely localized on a single site, whereas $\frac{1}{N}$ implies that the state is fully delocalized over the entire chain.

In order to obtain richer phase diagrams, we slightly modify our original chain, such that one of the sublattices can be shifted away from the symmetric zigzag geometry in Fig.~\ref{draft fig 1}(a) by $\Delta x$ and $\Delta y$ in $x$ and $y$ directions, respectively [see Fig.~\ref{draft fig 2}(a)].
In Fig.~\ref{draft fig 2}(b), we show the winding number $\nu$ and the localization $\mathrm{Loc}$ as functions of the polarization angle $\varphi$ and the shift in $y$ direction for $\Delta x=0$. 
The winding number shows large regions with nontrivial bulk topology, always with a phase transition at $\varphi=0$.
However, inspection of $\mathrm{Loc}$ reveals that strongly localized states are only present in a much smaller patch around $\Delta y \lesssim 0$ and $\varphi <0$. Less localized states (indicating a smaller band gap) are present in the region $\varphi>0$ and $\Delta y>0$. 
Most importantly, however, there are large regions where edge states predicted by the winding number are actually completely missing 
in the localization.
Note that winding numbers equal to $-1$ are numerical artifacts occurring for parameters where the Bloch bands are degenerate and, therefore, the winding number is ill-defined.
This happens typically (but not exclusively) at topological phase transitions.

Figure~\ref{draft fig 2}(c) shows the corresponding results when changing $\Delta x$ instead of $\Delta y$.
Also, here we observe large regions with nontrivial winding numbers that are only partially covered by actually localized states. 
Different from before, there are two small regions where the winding number takes values equal to $2$ (for $\Delta x \lesssim0.5a$ and $\varphi >0$) and $-1$ (for $\Delta x\gtrsim -0.5a$ and $\varphi<0$), respectively.
These are actual topological phases (and no numerical artifacts), however, without entailing edge states.

Figure~\ref{draft fig 2}(d) presents our numerical results for simultaneously varying $\Delta x$ and $\Delta y$.
Again, we obtain a rich phase diagram for the winding number, which is only partially confirmed by the localization. The few thin strips with winding numbers $-1$ and $3$ are  again  numerical artifacts, as the Bloch bands are degenerate at these parameters.
More interestingly, there is a relatively large region with winding number equal to 2 at the bottom right corner, which actually shows significantly larger localization values than the delocalized states nearby, indicating that there are indeed edge states present.
By explicitly computing the eigenstates for these parameters, one finds that there are indeed four states localized at the boundaries in accordance with the higher-order bulk topology. 
We conclude that, although weakening the bulk-boundary correspondence, the presence of long-range couplings can also enable higher-order topological phases.

\textit{Polarization as synthetic dimension.}
\begin{figure*}
    \centering
    \includegraphics[width=1\linewidth]{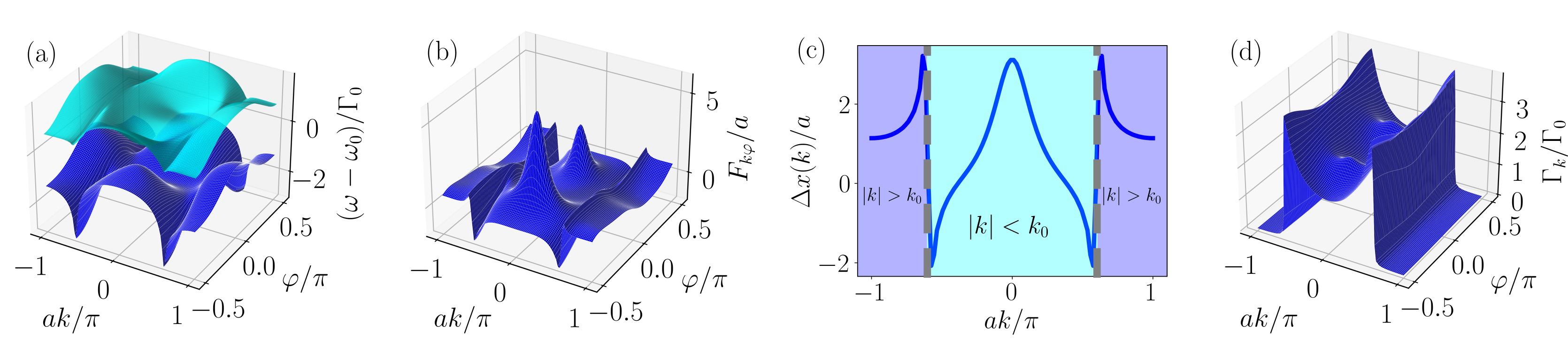}
    \caption{(a) Band structure $\omega=\omega_\pm(k,\varphi)$ of the Bloch Hamiltonian $\mathcal{H}_\mathrm{RM}(k,\varphi)$.
    (b) Berry curvature $F_{k\varphi}^-$ of the lower Bloch band.
    (c) Displacement after one pumping period as a function of $k$ [cf. Eq.~\eqref{eq draft displacement}].
    The cyan (purple) area indicates states inside (outside) the light cone $|k|=k_0$ (gray dashed lines at $ak/\pi=\pm0.6$).
    (d) Collective decay rates $\Gamma_k(\varphi)$ computed from Eq.~\eqref{eq draft collective decay rates} for the lower band.
    All plots used $N=50$, $a=0.3\lambda_0$, and $\Delta_0=\Gamma_0$.
    }
    \label{draft fig 3}
\end{figure*}
Its inherent $\pi$-periodicity allows us to consider the polarization angle as a synthetic $k$-space dimension of our otherwise one-dimensional system.
Synthetic dimensions enable studying higher-dimensional topological phases in low-dimensional systems, such as the two-dimensional quantum Hall effect in one-dimensional chains \cite{Lohse2016} or the four-dimensional quantum Hall effect in two- \cite{Kraus2013} or three-dimensional lattices \cite{Price2015}.
Typical examples of synthetic dimensions are periodic modulations of the lattice \cite{Lohse2016, Kraus2013}, internal atomic degrees of freedom \cite{Price2015}, or phase bias in Josephson junctions \cite{Klees2020, Klees2021}.
It has recently been shown that the bulk topology of such systems can be probed not only via spatial edge states but also using the time boundary effect \cite{Xu2025}.

Here, we show how the polarization angle can be used to simulate a two-dimensional Chern insulator with our original one-dimensional chain shown in Fig.~\ref{draft fig 1}(a) with dipole-dipole Hamiltonian $H(\varphi)$ in Eq.~\eqref{eq draft Hamiltonian}.
Our reasoning hereby follows arguments used in topological pumping protocols \cite{Kraus2013, Lohse2016, Svendsen2024}.
In order for our system to provide finite Chern numbers, we have to break time-reversal symmetry \cite{Ryu2010}, which can be achieved by an additional staggered on-site potential 
\begin{equation}
    H_\Delta(\varphi) = \Delta(\varphi) \sum_{i=1}^N (-1)^i \sigma_i^\dagger \sigma_i,
\end{equation}
with $\Delta(\varphi) = \Delta_0\cos(2\varphi)$, transforming our SSH model into a Rice-Mele model \cite{Rice1982}.
We note that the expression $\Delta(\varphi)$ should not imply that the on-site potential physically depends on the polarization angle, but rather that it is varied simultaneously with $\varphi$. 
For example, such an on-site potential could be realized using an inhomogeneous electric field (Stark shift) that is continuously varied while rotating the dipoles.

With the full Hamiltonian now being $H_{\rm RM}(\varphi) = H(\varphi) + H_\Delta(\varphi)$, we write the corresponding Bloch Hamiltonian as $\mathcal{H}_{\rm RM}(k,\varphi)$ with eigenstates $\ket{u_\pm(k,\varphi)}$ following from the Schrödinger equation $(\mathcal{H}_{\rm RM}(k,\varphi) - \omega_\pm(k,\varphi))\ket{u_\pm(k,\varphi)}=0$.
The energy bands $\omega_\pm(k,\varphi)$ are shown in Fig.~\ref{draft fig 3}(a), where one can see that the degeneracies at $(k, \varphi)=(\pm\pi/a, 0)$ and $(k, \varphi)=(\pm\pi/a, \pm\pi/2)$ are lifted by the on-site potential $\Delta(\varphi)$ such that the bands are well separated.
This enables defining the Berry curvature as $F_{k\varphi}^\pm = -2 \, \mathrm{Im}\braket{\partial_k u_\pm | \partial_\varphi u_\pm}$ \cite{Berry1984}, which is shown in Fig.~\ref{draft fig 3}(b) for the lower band $\ket{u_-}$.
(The upper band yields similar results.)
Integrating the Berry curvature over the synthetic Brillouin zone yields Chern numbers
\begin{equation}
    c_\pm = \frac{1}{2\pi} \int_{-\pi/a}^{\pi/a} \int_{-\pi/2}^{\pi/2} F_{k\varphi}^\pm(k,\varphi) \dd{\varphi}\dd{k}
    = \pm 1,
\end{equation}
showing that our system  indeed resembles a synthetic Chern insulator. Note that since we are working in the single excitation subspace, our Chern bands are not half filled (which would be necessary for an actual insulator), which is why edge state transport in our case is not quantized.
The real-space displacement $\Delta x(k)$ of the center of the excitation in state $\ket{u_\pm(k,\varphi)}$ after a full $\pi$-rotation of the polarization is given by \cite{Ke2020, Svendsen2024}
\begin{equation}\label{eq draft displacement}
    \Delta x(k) = \int_{-\pi/2}^{\pi/2} F_{k\varphi}^\pm (k,\varphi) \dd{\varphi},
\end{equation}
which is shown in Fig.~\ref{draft fig 3}(c).
We observe that, depending on the value of $k$, the excitation can be pumped up to $\approx2$ unit cells to the left or right, with a sharp transition at the so-called light line $|k|=k_0$, which has been shown to be a consequence of long-range couplings \cite{Svendsen2024}.
For actual transport of excitation through the chain, the states beyond the light line are most suitable since they are naturally subradiant, as shown in Fig.~\ref{draft fig 3}(d):
For $|k|>k_0$, the collective decay rates, which are given by 
\begin{equation}\label{eq draft collective decay rates}
    \Gamma_k(\varphi)^\pm = -\braket{\mathscr{L}[\ket{u_\pm(k,\varphi)}\bra{u_\pm(k,\varphi)}]},
\end{equation}
where $\mathscr{L}$ is the Lindblad operator in Eq.~\eqref{eq draft lindblad}, drop to zero, implying that excitation could be transported through the chain without decaying.

\textit{Perspectives.}
We have shown that the topological phases in zigzag-shaped SSH chains made of quantum emitters depend strongly on the polarization of the chain.
Depending on the exact choice of platform, variable polarization could be implemented experimentally either via the driving optical field or via other external fields controlling the dipole moment orientation.
For a situation similar to Fig.~\ref{draft fig 1}(b), where one needs to have at least two polarization orientations determining the topological phases, it would be sufficient if each emitter provides one ground state and at least two (ideally degenerate) excited states with perpendicular transition dipole moments. 
An example would be a transition from an $\ket{s}$ orbital to a $\ket{p}$ orbital in a hydrogenlike emitter.
By using linearly polarized light with polarization vector $\Vec{E}\propto (1, \ \pm 1, \ 0)^\mathrm{T}$, one can selectively drive the transition with dipole moment parallel to $\Vec{E}$, realizing the polarizations shown in Fig.~\ref{draft fig 1}(b).
Such a scheme has been successfully applied to topological plasmonic chains \cite{Poddubny2014, Sinev2015, Slobozhanyuk2015, Kruk2017, Kruk2019, Zhang2021, Moritake2022, Wu2024} and
should be applicable to atomic and atomlike emitters, like quantum dots, as well as color centers in diamond, as they can also provide multiple orthogonal transitions \cite{Alegre2007, Lamba2024}.
For the use of polarization as synthetic dimension, the required full $\pi$-rotation of the dipoles is more challenging. 
This could be realized using rotating molecular emitters in tweezer arrays, as the molecular transition dipole moments are rigidly aligned with the molecular axis \cite{Diehl2014}.
We believe that our findings can be helpful in settings in which one likes to study different topological phases without having to change the geometry.

\begin{acknowledgments}
We thank B. Trauzettel, S. Klembt, B. Olmos, and M. B. M. Svendsen for fruitful
discussions. 
This work was supported by
the German Science Foundation (Deutsche Forschungsgemeinschaft, DFG) in
the framework of  the Cluster of Excellence on Complexity and Topology
in Quantum Matter ct.qmat (EXC 2147, Project No.
390858490). A.P. gratefully acknowledges the Heisenberg Program of the
DFG  (Project PA 2508/3-1).
\end{acknowledgments}


\providecommand{\noopsort}[1]{}\providecommand{\singleletter}[1]{#1}%

\end{document}